\documentstyle[agupp,epsfig]{article}

\lefthead{MARTON ET AL.}
\righthead{PEROVSKITE EOS AND LOWER MANTLE COMPOSITION}
\received{March 29, 2000}
\revised{September 28, 2000}
\accepted{December 5, 2000}
\paperid{2000JB900457}
\cpright{AGU}{2001}
\ccc{0148-0227/01/2000JB900457\$09.00}
\authoraddr{R.~E.\/ Cohen, and F.~C.\/ Marton,
Geophysical Laboratory and Center for High Pressure Research,
Carnegie Institution of Washington,
5251 Broad Branch Road, NW, Washington, D. C. 20015-1305.
(cohen@gl.ciw.edu; marton@gl.ciw.edu)}

\authoraddr{J.\/ Ita, Shell International Exploration and
Production, P. O. Box 60, Rijswijk, 2280 AB, Netherlands.
(j.j.ita@siep.shell.com)}

\begin{document}

%% ------------------------------------------------------ %%
%
%%  TITLE
%
%% ------------------------------------------------------ %%

\title{Pressure-volume-temperature equation of state \\
of MgSiO$_{3}$ perovskite from molecular dynamics \\
and constraints on lower mantle composition}

  % Type the title of your manuscript between the
  % curly brackets in a \title command.  Capitalize
  % only acronyms, first letter of the first word,
  % first letter of proper nouns, first letter of
  % the first word after colons, and first letter
  % of the first word of a subtitle.  If the title
  % exceeds one line, break it so that the first line
  % is longer than the second line; break the title
  % before articles, prepositions, and conjunctions.
  % To break the title, type a double backslash where
  % you want the break to occur, as shown above.

%% ------------------------------------------------------ %%
%
%%  AUTHOR NAMES, AFFILIATIONS, and ALTERNATE AFFILIATIONS
%
%% ------------------------------------------------------ %%

\author{Frederic C.\/ Marton, Joel Ita,\altaffilmark{1} and Ronald
E. Cohen}
\affil{Geophysical Laboratory and Center for High Pressure Research,
\\Carnegie Institution of Washington, Washington, D. C.}

\altaffiltext{1}{Now at Shell International Exploration and
Production, Rijswijk, Netherlands.}

\begin{abstract}
The composition of the lower mantle can be investigated
by examining densities and seismic velocities of
compositional models as functions of depth.  
In order to do this it is necessary to know the 
volumes and thermoelastic properties of the compositional
constituents under lower mantle conditions.
We determined the thermal equation of state (EOS)
of MgSiO$_{3}$ perovskite using the nonempirical 
variational induced breathing (VIB) interatomic potential
with molecular dynamics simulations
at pressures and temperatures of the lower mantle.
We fit our pressure-volume-temperature results to a thermal EOS
of the form
$P(V,T)$ = $P_{0}(V,T_{0})$ + $\Delta P_{\rm th}(T)$, where
$T_{0}$ = 300 K and $P_{0}$ is the isothermal Universal EOS.
The thermal pressure $\Delta P_{\rm th}$ can be represented
by a linear relationship $\Delta P_{\rm th} = a + b T$.
We find $V_{0}$ = 165.40 \AA$^{3}$, $K_{T_{0}}$
= 273 GPa, $K^{\prime}_{T_{0}}$ = 3.86,
$a$ = -1.99 GPa, and $b$ = 0.00664 GPa K$^{-1}$
for pressures of 0-140 GPa and temperatures of 300-3000 K.
By fixing $V_{0}$ to the experimentally determined value
of 162.49 \AA$^{3}$ and
calculating density and bulk sound velocity profiles
along a lower mantle geotherm
we find that the lower mantle cannot consist solely of
(Mg,Fe)SiO$_{3}$ perovskite with X$_{\rm Mg}$ ranging from
0.9-1.0.  Using pyrolitic compositions of 67 vol \%
perovskite (X$_{\rm Mg}$ = 0.93-0.96) and 33 vol \% magnesiow\"ustite
(X$_{\rm Mg}$ = 0.82-0.86), however, we obtained density and
velocity profiles that are in excellent agreement
with seismological models for a reasonable geotherm.
\end{abstract}

\begin{article}
\section{Introduction}
\begin{figure}
\centerline{\epsfig{file=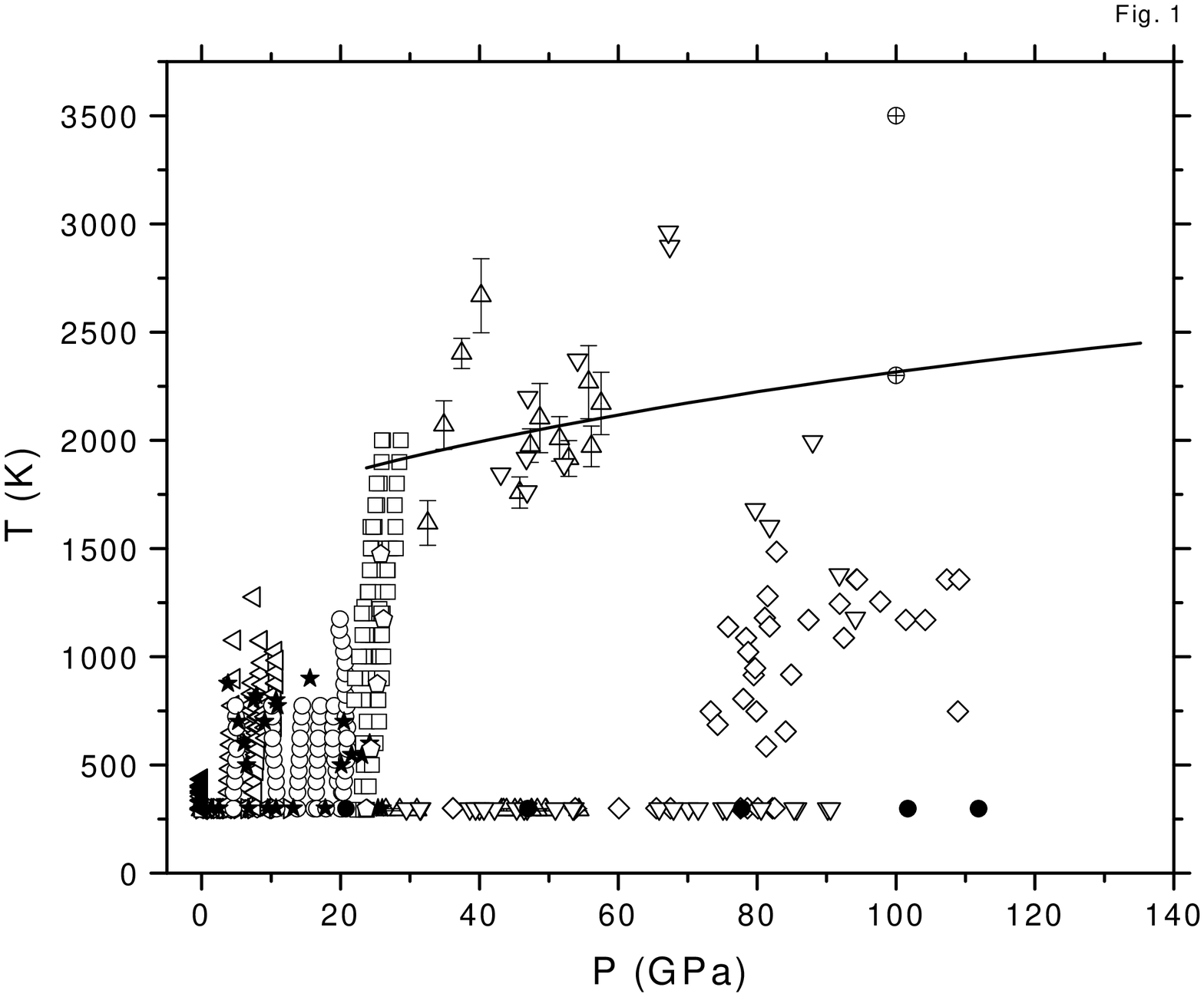,width=5in}}
\caption{Experimental pressure-temperature coverage of
(Mg,Fe)SiO$_{3}$ perovskite.  Open symbols are for static
experiments on X$_{\rm Mg}$ = 1.0, solid symbols
are for X$_{\rm Mg}$ = 0.9, and
open symbols with crosses are shock wave data.
Hexagons, {\it Kudoh et al.,} [1987];
right-facing triangles, {\it Ross and Hazen,} [1990]; 
left-facing triangles, {\it Wang et al.,} [1994];
upward pointing triangles, {\it Fiquet et al.,} [1998];
downward pointing triangles, {\it Fiquet et al.,} [2000];
diamonds, {\it Saxena et al.,} [1999];
open circles, {\it Utsumi et al.,} [1995];
solid circles, {\it Knittle and Jeanloz,} [1987];
squares, {\it Funamori et al.,} [1996];
stars, {\it Mao et al.,} [1991];
pentagons, {\it Kato et al.,} [1995];
circle with cross, {\it Duffy and Ahrens,} [1993].
Solid curve is the lower mantle geothermal gradient of
{\it Brown and Shankland} [1981]. 
Static experimental work has direct
coverage of approximately the upper one third.  The shock wave 
data point that falls near the geotherm is perovskite plus
magnesiow\"ustite.}
\label{fig1}
\end{figure}

By comparing density and seismic velocity profiles of
compositional models with seismological models it is possible
to investigate the composition of the Earth's lower mantle.
Therefore an understanding of the high pressure, high temperature
properties, and equation of state (EOS) of
(Mg,Fe)SiO$_{3}$ perovskite is vital to our understanding
of the lower mantle, as this mineral accounts for
perhaps two thirds of the mineralogy of this region \cite{bina98}
and one third of the volume of the entire
planet.  Experiments have been performed
up to pressures approaching those found at the base of the mantle,
but direct coverage of the lower mantle geotherm has been limited
to perhaps the uppermost one third (\callout{Figure~\ref{fig1}}).
Experiments include resistively heated multianvil apparati 
\cite{wang94, kato95, utsumi95, funamori96}, 
diamond anvil cells (DACs) at ambient temperature
\cite{kudoh87, ross90},
resistively heated DACs \cite{mao91, saxena99},
laser heated DACs \cite{knittle87,fiquet00}, and
shock wave experiments \cite{duffy93, akins99}.

Theoretical methods, such as those used here, permit the 
investigation of thermodynamic and thermoelastic properties of
minerals at high pressures and temperatures.
Previous theoretical work on orthorhombic
MgSiO$_{3}$ perovskite (Mg-pv) 
has been done using molecular dynamics (MD) using the 
semi\-empirical potential of {\it Matsui} [1988]
at 0-1000 K and 0-10 GPa (at 300 K) \cite{matsui88} and 300-5500 K
and 0-400 GPa \cite{belonoshko94, belonoshko96} and combined 
with lattice dynamics at 500-3000 K and 0-100 GPa \cite{patel96}.
{\it Wolf and Bukowinski} [1987] used a rigid ion
modified electron gas (MEG) model at 0-800 K and 0-150 GPa,
while {\it Hemley et al.} [1987] used a MEG model combined
with shell stabilized ion charge densities at 0-2500 K and
0-200 GPa.  {\it Cohen} [1987] did quasi-harmonic
lattice dynamics calculations from 0-3000 K
and 0-150 GPa using the potential induced breathing (PIB) 
model.  First principles static lattice calculations ($T$ = 0 K)
have also been done up to pressures of 150 GPa, using plane wave
pseudopotential (PWPP) MD \cite{wentzcovitch95} and the
linearized augmented plane wave (LAPW) method \cite {stixrude93}.
{\it Karki et al.} [1997] also used the PWPP method to
examine the athermal elastic moduli of Mg-pv.

In addition to the equation of state the thermodynamic
stability of perovskite in the lower mantle is an open
question and has been studied with experimental
and theoretical methods.  Work by {\it Meade et al.} [1995],
{\it Saxena et al.} [1996, 1998], and
{\it Dubrovinsky et al.} [1998]
indicate that perovskite will break down to oxides under
lower mantle conditions.  
The thermodynamic analysis of {\it Stixrude and
Bukowinski} [1992] and the experimental 
work of {\it Serghiou et al.} [1998],
however, suggest that it will not.

We used MD simulations using the nonempirical variational
induced breathing (VIB) model, similar to the model of
{\it Wolf and Bukowinski} [1988],
to investigate the properties and EOS of Mg-pv
at pressures (0-140 GPa) and temperatures (300-3000 K)
that cover the bulk of the conditions found in the lower mantle.
Newton's equations of motion are solved as functions of time,
and equilibrium properties are obtained as time averages
over a sufficiently long interval.  
This accounts for all orders of anharmonicity but not for
quantum lattice dynamics effects.  Thus our results are more
appropriate at high temperatures above the Debye temperature
(1076 K \cite{anderson98}) and are entirely suitable for the Earth's mantle.
We then used our results, combining them with data for other
components and phases where appropriate, to calculate density and
seismic velocity profiles along a geothermal gradient for different
compositional models.  These profiles were then compared with 
profiles from a reference Earth model in order to test the 
compositional models' fitness for the lower mantle.

\section{Method}
MD simulations were performed on
supercells of 540-2500 atoms of orthorhombic ({\it Pbnm})
Mg-pv using the nonempirical VIB potential.
VIB is a Gordon-Kim type model \cite{gordon72},
where the potential is obtained by overlapping 
ionic charge densities which are computed using the local
density approximation (LDA) \cite{hedin71}.  The total energy is a sum of
(1) the self-energy of each atom, (2) the long-range 
electrostatic energy computed using the Ewald method, and
(3) the short-range interaction energy, i.e.,
the sum of the kinetic, short-range electrostatic,
and exchange-correlation overlap energies from the LDA.
Free oxygen ions are not stable and are stabilized 
in VIB by surrounding
them with Watson spheres with charges equal in magnitude but opposite
in sign to the ions, e.g., 2+ spheres for O$^{2-}$.  
Interactions are obtained for overlapping ion pairs at different
distances with different Watson sphere radii for each oxygen.
The interactions are fit with a 23-parameter analytical
expression as functions of the interatomic distance $r$ and
$U_{i} = z_{i} / R_{i}$, where $U_{i}$, $z_{i}$, and $R_{i}$ 
are the Watson sphere potential, charge, and radius for atom $i$.
For each oxygen $i$, $R_{i}$ is adjusted to minimize the total
energy at each time step.
This gives an effective many-body potential.
The oxygen ions respond to changes in their environment.
For example, they are compressed at high pressures relative
to low pressures.
Previous work on Mg-pv using 
the related PIB model
gave an equation of state in excellent agreement with 
experiment \cite{cohen87}.  Also, work using VIB
on MgO has shown that this model accurately predicts
EOS and thermal properties \cite{inbar95}.

\begin{figure}
\centerline{\epsfig{file=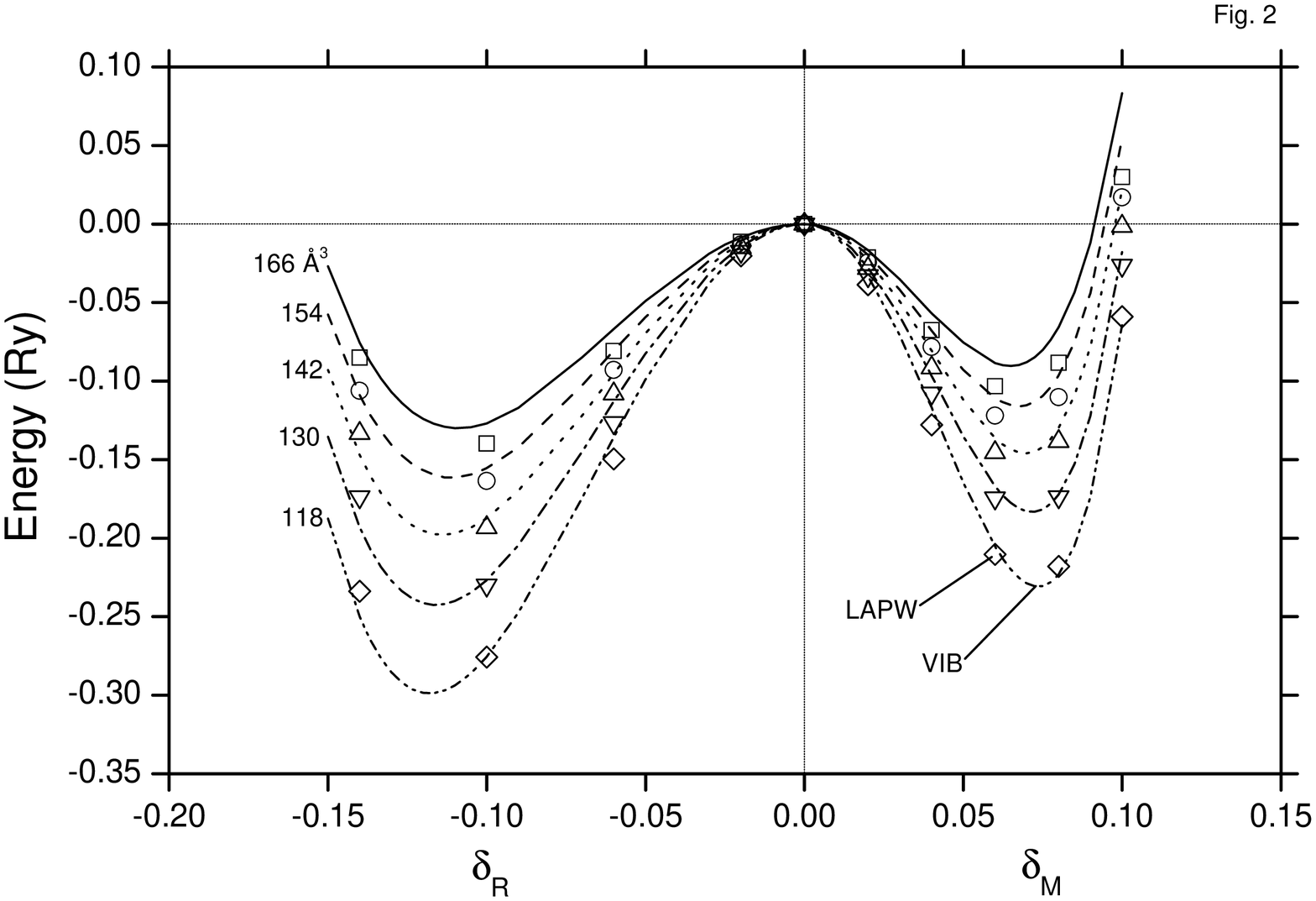,width=5in}}
\caption{Energetics of octahedral rotations:
calculated total energies relative to the cubic structure
($Pm$3$m$) using VIB (curves) and LAPW [{\it Stixrude and
Cohen,} 1993; {\it Hemley and Cohen,} 1996]
(symbols) as a function of (left) $R$ and (right) $M$ point
rotation angles as represented by the fractional change
in the oxygen coordinate ($\delta_{R}$ and $\delta_{M}$).
The orthorhombic structure ({\it Pbnm}) 
occurs at $\delta_{R}$ = 0.0912 and $\delta_{M}$ = 0.0766
at $P$ = 0 GPa.}
\label{fig2}
\end{figure}

Nominally charged ions give semiquantitative, but
less accurate, results than desired for Mg-pv.
This is due to a small degree of covalent bonding relative
to ionic bonding as revealed by electronic structure
calculations of cubic ($Pm$3$m$) Mg-pv 
performed using the first principles LAPW method \cite{cohen89}.
These calculations show that while the Mg is nearly a perfectly
spherical Mg$^{2+}$ ion, there is some charge transfer from
O to Si and a small covalent bond charge 
(involving $\ll$ 0.1 electrons) between Si and O.
By varying the ionic charges on Si (3.1+ to 3.55+) and
O (1.70- to 1.85-) to account for the covalency,
but otherwise using the same methods as before, 
we compared the resulting zero temperature pressure-volume data to
the LAPW results of 
{\it Stixrude and Cohen} [1993].  The best agreement
was found with Si$^{3.4+}$ and O$^{1.8-}$.
Using these charges, the resulting potential gives
excellent agreement with the octahedral rotation energetics
obtained using the LAPW method \cite{stixrude93, hemley96}
(\callout{Figure~\ref{fig2}}).

MD runs were performed for 20 ps with a 1 fs time step
using a sixth-order Gear predictor-corrector scheme
\cite{gear71} in the constant pressure-tempera\-ture ensemble
using the thermostat and barostat of {\it Martyna et al.} [1994].
Initial atomic positions for the genesis run were the same 
as those for the unrotated octahedra of
{\it Stixrude and Cohen} [1993].  
Subsequent runs at higher pressures or temperatures
used the positions generated by a previous run.
Statistical ensembles were obtained in $\sim$2000 iterations
for the genesis run and in $<$ 1000 iterations for subsequent runs.

We fitted pressure-volume-temperature (P-V-T) 
data obtained from the MD simulations
to a thermal EOS of the form
\begin{equation}
   P(V,T) = P_{0}(V,T_{0}) + \Delta P_{\rm th}(T)
\end{equation}
with a reference temperature $T_{0}$ = 300 K
and the thermal pressure $\Delta P_{\rm th}$ relative to the 300 K
isotherm.  We analyzed our results using the third-order
Birch-Murnaghan isothermal EOS \cite{birch52}
\begin{equation}
   P_{0} = 3 K_{T_{0}} f (1 + 2f)^{5/2} (1 - \xi f),
\end{equation}
with the Eulerian strain variable $f$ and the coefficient
$\xi$ given by
\begin{equation}
   f \equiv \frac{1}{2} \left[ \left( \frac {V_{0}}{V} \right)
   ^{2/3} - 1 \right] , ~~~~
   \xi \equiv - \frac{3}{2} ( K^{\prime}_{T_{0}} - 4 ),
\end{equation}
and the Universal EOS \cite{vinet87}
\begin{equation}
   P_{0} = 3 K_{T_{0}} (1 - x) \, x^{-2} \exp \left[ 
   \frac {3}{2} \, (K^{\prime}_{T_{0}} - 1) (1 - x) \right],
\end{equation}
where $x = ( V / V_{0} )^{1/3}$,
$K_{T}$ is the isothermal bulk modulus, and $K^{\prime}_{T}$
is its pressure derivative.

$\Delta P_{\rm th}$ is given by
\begin{equation}
   \Delta P_{\rm th} = \int^{T}_{T_{0}} \left( \frac {\partial {P}}
   {\partial T} \right)_{V} d{\hat{T}} =
   \int^{T}_{T_{0}} ( \alpha K_{T} ) d{\hat{T}},
\end{equation}
where $\alpha$ is the volume coefficient of thermal expansion.
If $\alpha K_{T}$ is independent of T, then the thermal
pressure can be represented by a linear relationship
\cite{anderson80,anderson83}
\begin{equation}
   \Delta P_{\rm th} = a + b T ,
\end{equation}
which applies 
to a wide range of solids at high $T$, including minerals,
alkali metals, and noble gas solids \cite{anderson84}.
The linearity of $\Delta P_{\rm th}$
in $T$ starts at 300 K \cite{masuda95} in minerals.
There should be a small anharmonic correction at high $T$, which
results in an additional $c\,T^{2}$ term in (6),
which is often sufficiently small so that it can be neglected
\cite{anderson84}.  Indeed, including the $c\,T^{2}$ term
did not statistically improve our fits.  Likewise,
our fits were not improved by including volumetric compression terms 
[e.g., {\it Jackson and Rigden}, 1996].

Once the P-V-T data were fit, other parameters, such as
$\alpha$,
\begin{equation}
   \alpha = { \left( \frac {\partial {\:\ln V}}
                {\partial T} \right) }_{P},
\end{equation}
its volume dependence, the Anderson-Gr\"uneisen parameter,
\begin{equation}
   \delta_{T} = { \left( \frac {\partial {\:\ln \alpha}}
                {\partial {\:\ln V}} \right) }_{T},
\end{equation}
the Gr\"uneisen parameter,
\begin{equation}
   \gamma = \frac {\alpha K_{T} V}{C_{V}},
\end{equation}
and its volume dependence,
\begin{equation}
   q = { \left( \frac {\partial {\:\ln \gamma}}
       {\partial {\:\ln V}} \right) }_{T},
\end{equation}
were obtained by numerical differentiation.  $C_{P}$ was
found by calculating enthalpy ({\it U \rm + PV\,}) 
at each MD run point and
differentiating with respect to $T$.
Using the relation $C_{P} = C_{V} (1 + \gamma \alpha T)$,
equation (9) can be rewritten as
\begin{equation}
   \gamma = \left( \frac {C_{P}}{\alpha K_{T} V} -
            T \alpha \right)^{-1}.
\end{equation}

\section{Results}
\begin{figure}
\centerline{\epsfig{file=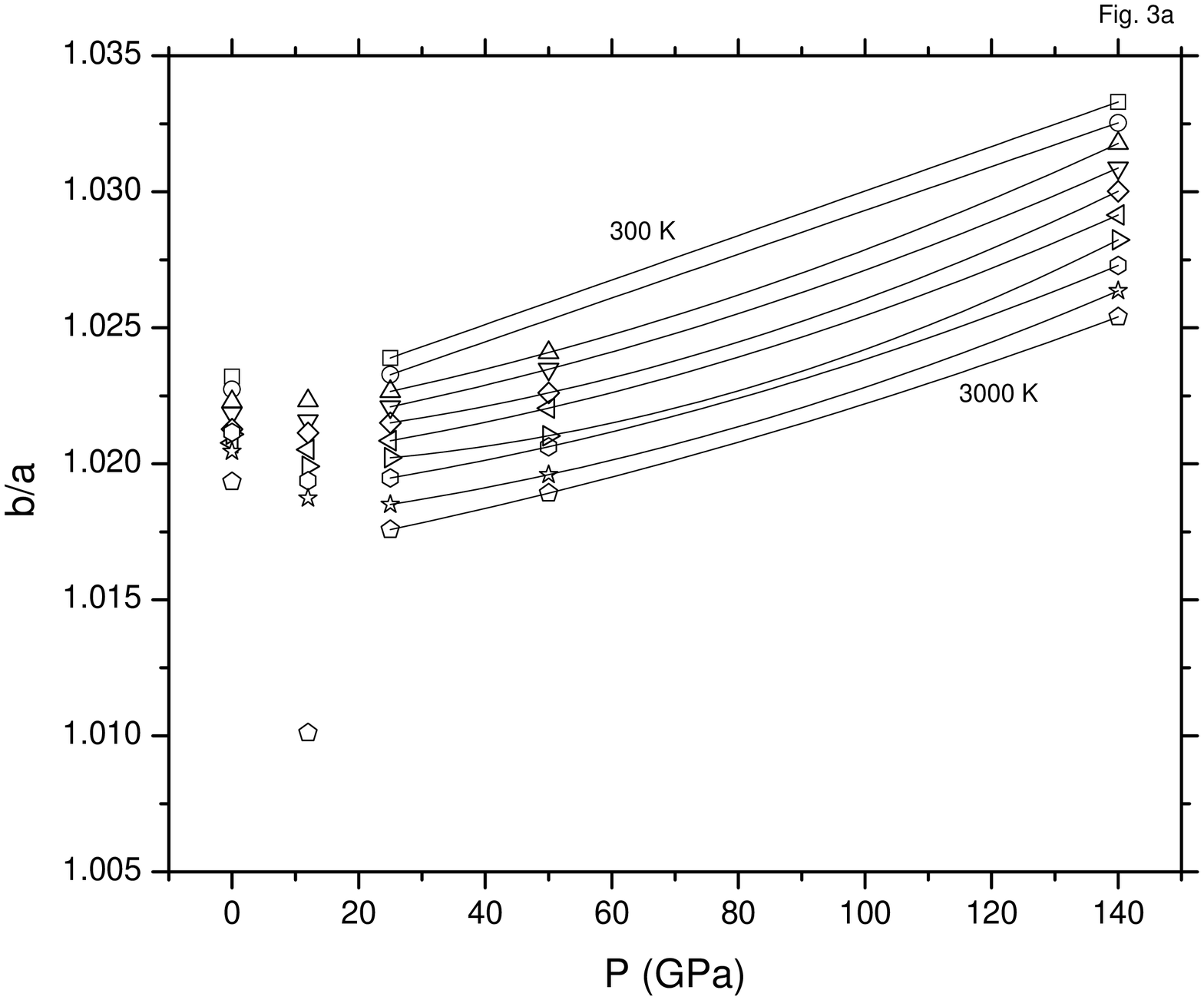,width=5in}}
\centerline{\epsfig{file=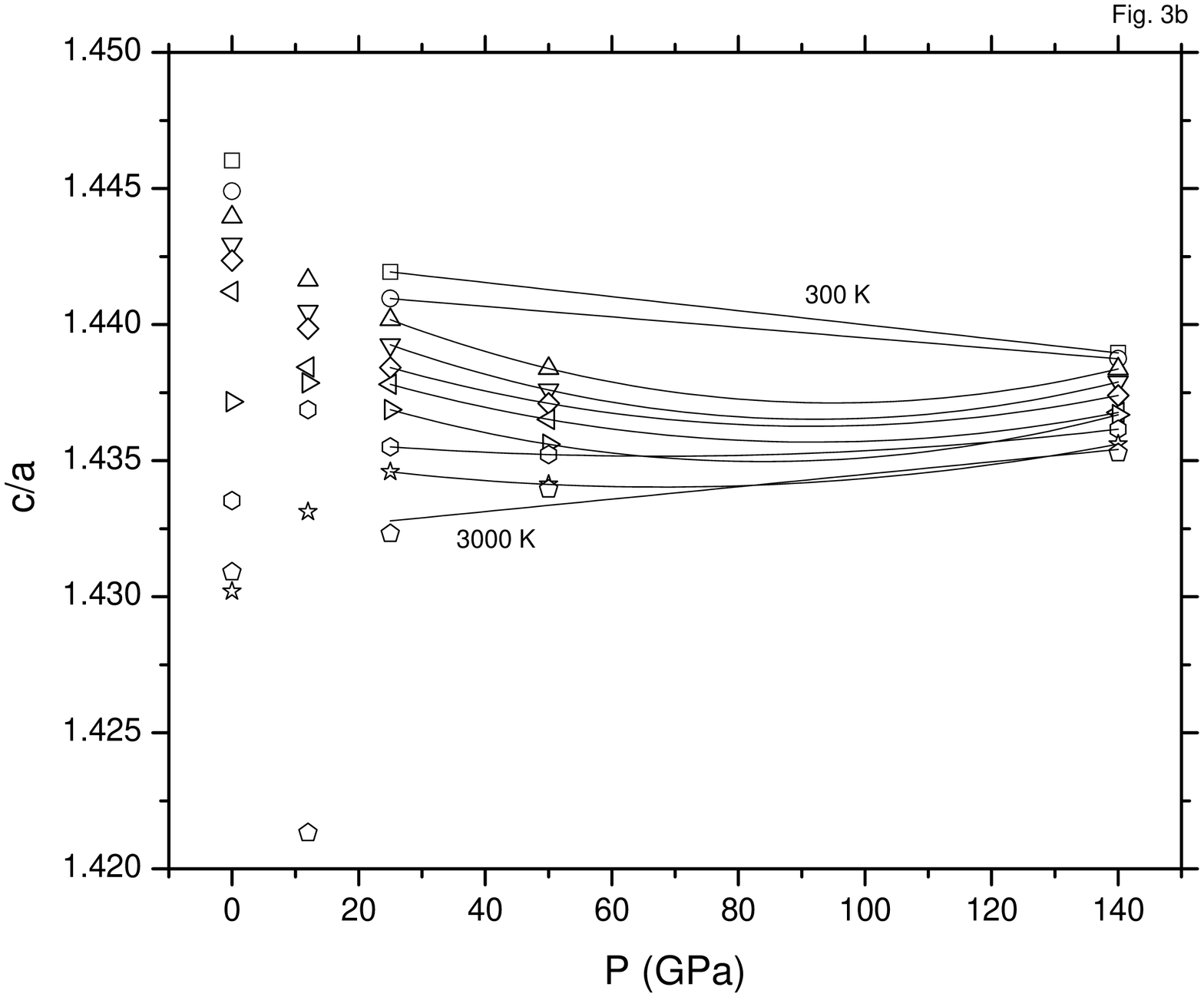,width=5in}}
\clearpage
\centerline{\epsfig{file=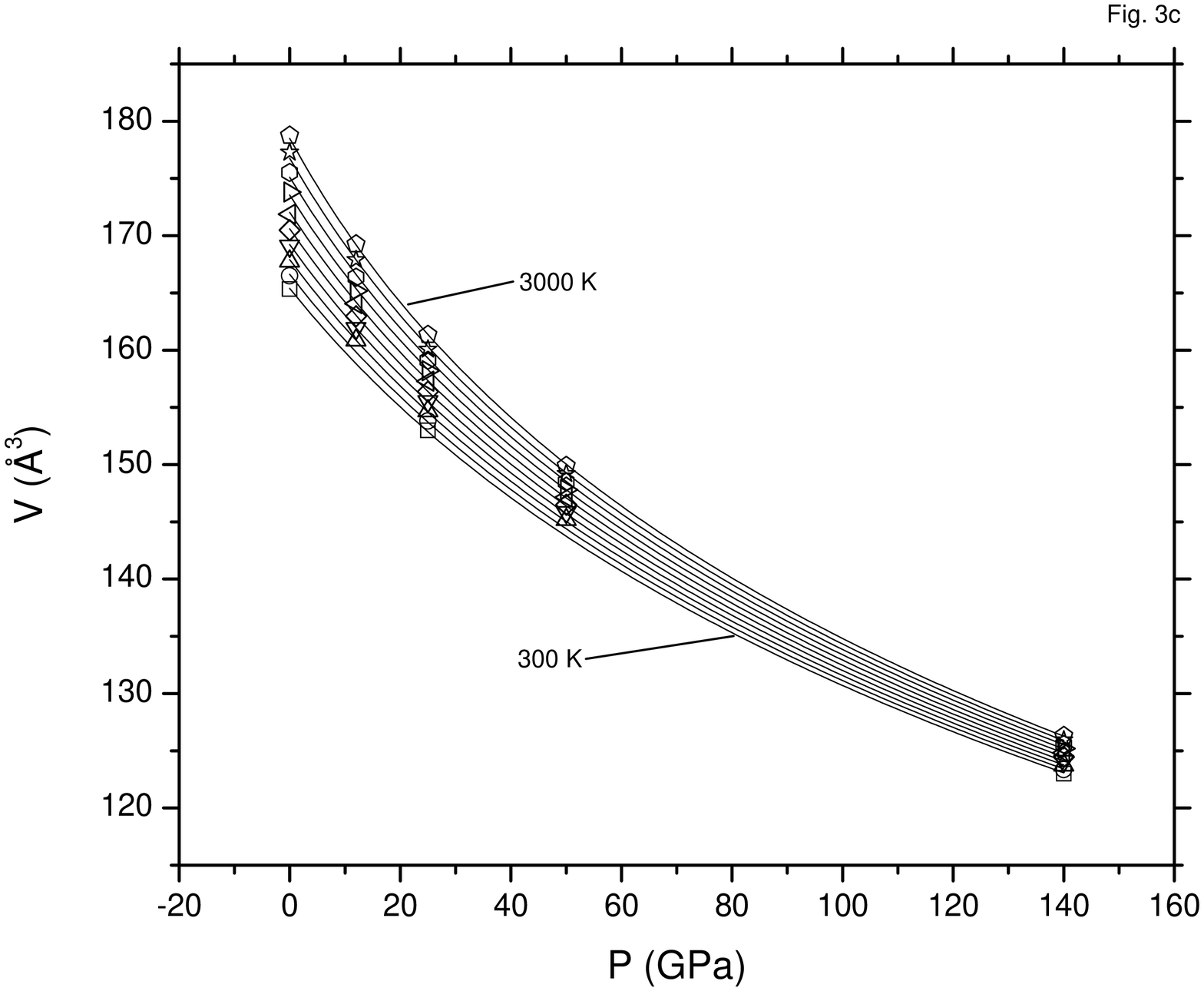,width=5in}}
\caption{Structural parameters and volumes of MD results:
(a) $b/a$ axis ratio and (b) $c/a$ axis ratio.
At low pressures and high temperatures the structure 
begins to deviate from orthorhombic {\it Pbnm}, but no
phase transitions, with the possibility of the 
structure heading toward cubic at 12 GPa and 3000 K.
(c) Volume.  Symbols are MD results and curves are Universal EOS
fits every 300 K.}
\label{fig3}
\end{figure}

MD calculations were performed at 46 pressure-temp\-er\-ature
(P-T) points, ranging from 0-140 GPa and 300-3000 K, thus covering the
P-T conditions of the lower mantle (\callout{Table~\ref{tbl1}}).  
Axial ratios $b/a$ and $c/a$ follow smooth trends at lower mantle
pressures (\callout{Figures~\ref{fig3}}a-3b), with deviations
away from the {\it Pbnm} structure seen at $P$ $\le$ 12 GPa
and $T$ $\ge$ 2400 K.
At 3000 K and 12 GPa the structure can be observed
to move towards cubic lattice parameters 
($b/a$ $\rightarrow$ 1, $c/a$ $\rightarrow$
$\sqrt{2}\,$).  This point,
in particular, is close to the melting curve of Mg-pv
\cite{heinz87,knittle89,poirier89}, and such temperature-induced
phase transitions have been predicted for Mg-pv by MD
\cite{wolf87} and lattice dynamics \cite{matsui91}.
In addition, it has been found for RbCaF$_{3}$
and KMnF$_{3}$ perovskites by MD simulations \cite{lewis89,nose89}
and observed experimentally in the fluoride perovskite
neighborite (NaMgF$_{3}$) \cite{chao61}.
Regardless of these changes, the volumes are well-behaved
as functions of pressure and temperature (Figure 3c),
and no other structural anomalies or melting was
encountered during runs at other P-T points.

The resulting P-V-T EOS
fits and their reduced $\chi^{2}$ values 
are given in \callout{Table~\ref{tbl2}}.  Both the 
Birch-Murnaghan and Universal EOSs are in excellent agreement
with experimental results.  As for the accuracy of our thermal
pressure expression, RMS differences between volumes found using
equation (1) for either EOS and those found using isothermal EOSs are of 
the order of 10$^{-3}$ \AA$^{3}$.  As for the differences between
the two equation of state forms, {\it Jeanloz} [1988] 
compared them over moderate
compressions and found that they are quite similar.  More recent
work by {\it Cohen et al.} [2000] supports this, but they found
that for large ($>$30\%) strains and for determining parameters such
as $V_{0},~ K_{T_{0}}, ~{\rm and } ~ K^{\prime}_{T_{0}}$, the
Universal EOS works best.  So, for the rest of our analyses
we use that here.  However, given that under the pressure range
studied, compressions will be no greater than 25\%, so we can
confidently compare our results with those of others that were
found with the Birch-Murnaghan EOS.

Once we determined the EOS parameters, we determined volumes
for Mg-pv over a wide P-T range
\linebreak
(-10-150 GPa, 0-3300 K).  Using this extended data set, we
determined the isothermal bulk modulus 
$[ K_{T} =  
%\linebreak 
-V ( {\partial P}/{\partial V} )_{T} ]$
at those points and
$\alpha$, $\delta_{T}$, $\gamma$, and $q$
using equations (7), (8), (11), and (10).
We found $({\partial K}/{\partial T} )_{P=0}$
= -0.0251 GPa K$^{-1}$, close to those found experimentally,
-0.023 GPa K$^{-1}$ \cite{wang94} and -0.027 GPa K$^{-1}$ \cite{fiquet98},
as well as one found by the inversion of multiple experimental
data sets, -0.021 GPa K$^{-1}$ \cite{jackson96}.

\begin{figure}
\centerline{\epsfig{file=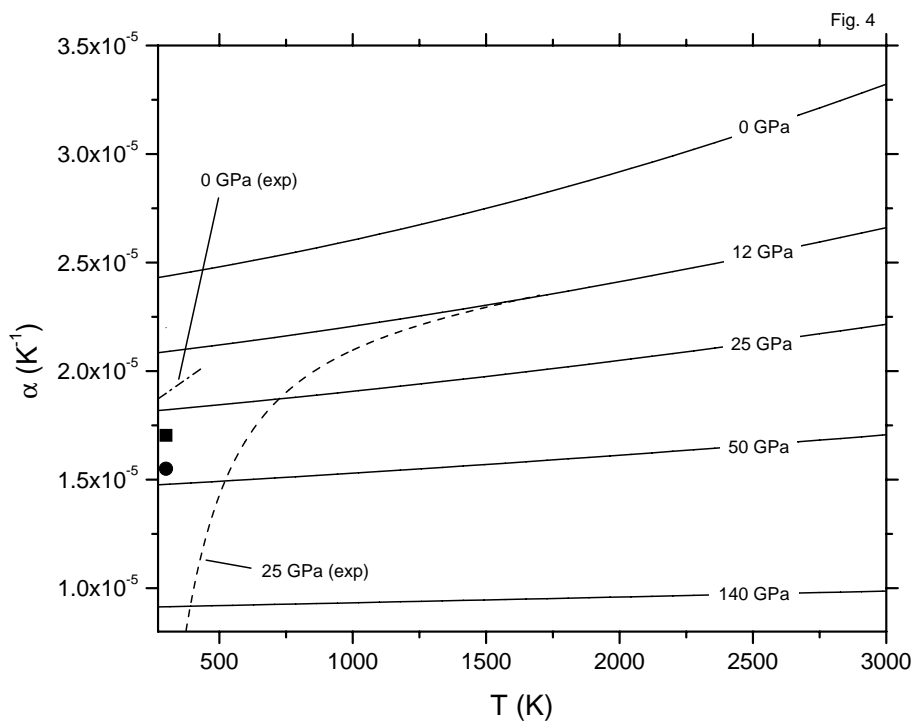,width=5in}}
\caption{Volume coefficient of thermal expansion.  Solid curves are
from this work.  Other symbols, from experimental work as follows:
dashed curve (0 GPa), {\it Wang et al.} [1994]; 
circle (0 GPa), {\it Fiquet et al.} [1998]; 
square (0 GPa) and dashed-dotted line (25 GPa),{\it Funamori et al.}
[1996]}
\label{fig4}
\end{figure}

\begin{figure}
\centerline{\epsfig{file=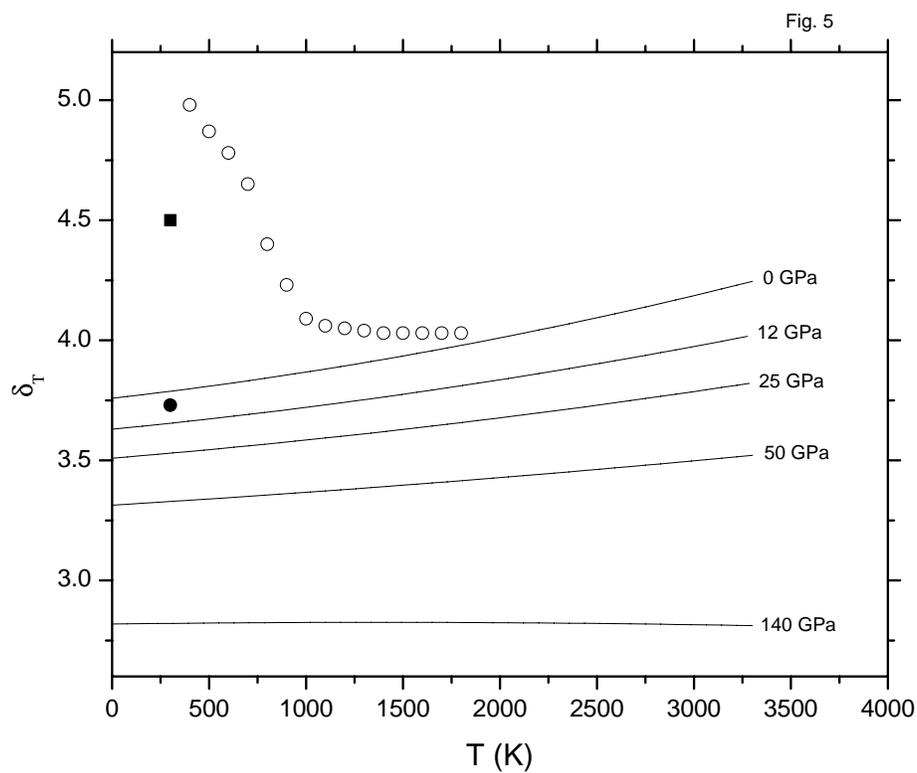,width=5in}}
\caption{Anderson-Gr\"uneisen parameter as a function of 
temperature.  The circle is the 
experimentally determined value
from {\it Wang et al.} [1994]. The square is the value
determined by {\it Masuda and Anderson} [1995] from the
experimental data of {\it Utsumi et al.} [1995].
Open circles are $\delta_{T}$s determined from Debye 
theory by {\it Anderson} [1998].}
\label{fig5}
\end{figure}

\callout{Figure~\ref{fig4}} shows $\alpha$ at pressures
from 0 to 140 GPa.  Experimental results shown have higher
$T$ slopes, as well as higher extrapolated values for most temperatures,
though an average value found by {\it Jackson and Rigden} [1996]
at zero GPa of $2.6 \times 10^{-5} \: {\rm K}^{-1}$ over 300-1660 K
matches our value over the same $T$ range exactly.
{\it Kato et al.} [1995] found an average value of 
$2.0 \pm 0.4 \times 10^{-5} \: {\rm K}^{-1}$ over 298-1473 K and 25 GPa, 
close to our average value of $1.89 \times 10^{-5} \: {\rm K}^{-1}$
at the same conditions.  The lack of quantum effects in our MD
results can be seen at low temperatures;
instead of tending toward zero at zero temperature,
$\alpha$ remains large.
The $\delta_{T}$ increases with $T$, but decreases
as $P$ increases (\callout{Figure~\ref{fig5}}).  The dependency
of $\delta_{T}$ on $T$ decreases as a function of $P$, with
convergence to a value of 2.87 at $\sim$130 GPa, similar to the 
high-pressure behavior in MgO found by the same MD method \cite{inbar95}.
Our value of $\delta_{T}$ = 3.79 at 
ambient conditions is in excellent agreement with the 
experimentally derived value of {\it Wang et al.} [1994] of 3.73,
though it is lower than the value of 4.5, found by {\it Masuda and
Anderson} [1995] from the experimental data of {\it Utsumi et al.}
[1995].
Other theoretical determinations of $\delta_{T}$ are much higher,
however.  {\it Hama and Suito} [1998a] found a value of
5.21, on the basis of calculations using the LAPW results of 
{\it Stixrude and Cohen} [1993] and
the Debye approximation for lattice vibration.
Using MD and semiempirical potentials \cite{matsui88}, 
{\it Belonoshko and Dubrovinsky} [1996] found $\delta_{T}$ = 5.80, and
{\it Patel et al.} [1996] found $\delta_{T}$ = 7.0 using
the same potentials, combining MD with lattice dynamics.
{\it Anderson} [1998], using Debye theory, calculated zero pressure
values ranging from 4.98 at 400 K
to $\sim$4 at 1000 K $\le$ $T$ $\le$ 1800 K, coming close to
our values at those temperatures.

\begin{figure}
\centerline{\epsfig{file=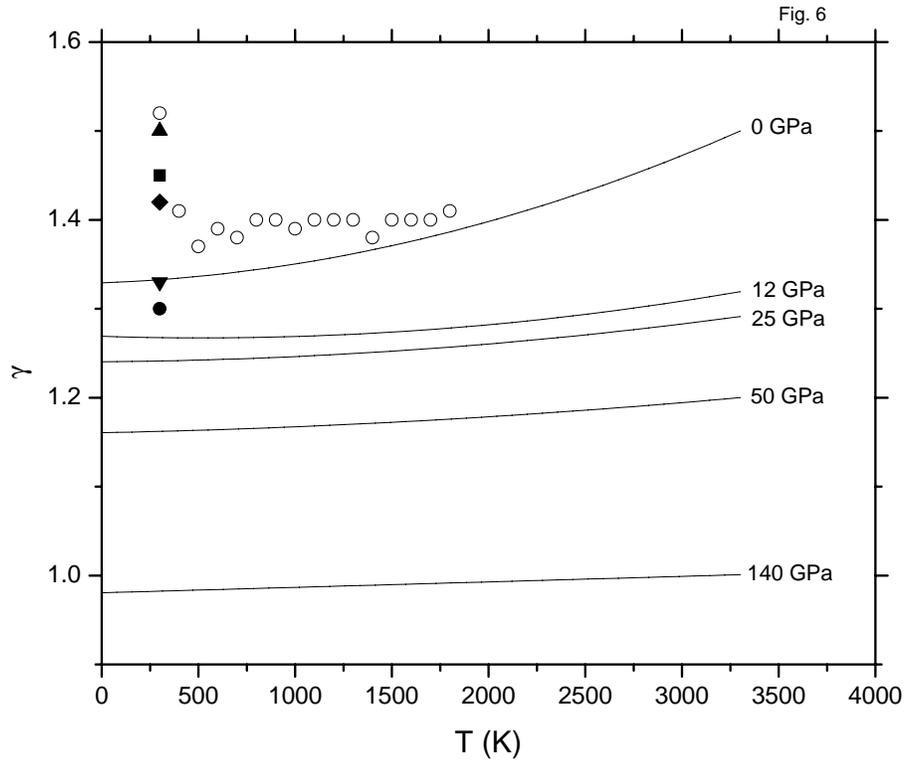,width=5in}}
\caption{Gr\"uneisen parameter as a function of
temperature.  The solid circle is the 
experimentally determined value from {\it Wang et al.} [1994],
and the square is the value determined by {\it Masuda and Anderson}
[1995] from the experimental data of {\it Utsumi et al.} [1995].
Other symbols are from inversion of multiple experimentally
determined data sets: triangle, {\it Bina} [1995];
inverted triangle, {\it Jackson and Rigden} [1996]); and 
diamond, {\it Shim and Duffy} 2000].  Open circles are
$\gamma$s determined from Debye theory by {\it Anderson} [1998].}
\label{fig6}
\end{figure}

The Gr\"uneisen parameter 
also increases as a function of $T$ and decreases as a
function of $P$ and approaches a value of 1 at
pressures of 130-140 GPa (\callout{Figure~\ref{fig6}}).
Our value of 1.33 at zero $P$ and 300 K matches 
well with {\it Wang et al.{\rm 's}} [1994] value of 1.3 and 
{\it Utsumi et al.{\rm 's}} [1995] and {\it Masuda and Anderson{\rm 's}}
[1995] value of 1.45.
{\it Stixrude et al.} [1992], using the experimental data of
{\it Mao et al.} [1991], found a higher value of
$\gamma$ = 1.96.  Values determined by
the inversion of multiple experimentally determined P-V-T data
sets match well also: 1.5 \cite{bina95}, 1.33 \cite{jackson96},
and 1.42 \cite{shim00}.
Two shock wave studies find a value of 1.5 \cite{duffy93, akins99}
on the basis of limited data sets: four for the former (with $q$ fixed equal
to 1) and two for the latter.
Values of $\gamma$ from theoretical studies range from very close
to ours, 1.279 \cite{hama98} and 1.44 \cite{hemley87}, 
to 1.97 \cite{belonoshko96, patel96}.
{\it Anderson{\rm 's}} [1998] zero pressure, 300 K value of 1.52 is
somewhat higher than ours, but at 400-1800 K his values of
$\sim$1.4 are very close to ours.

\begin{figure}
\centerline{\epsfig{file=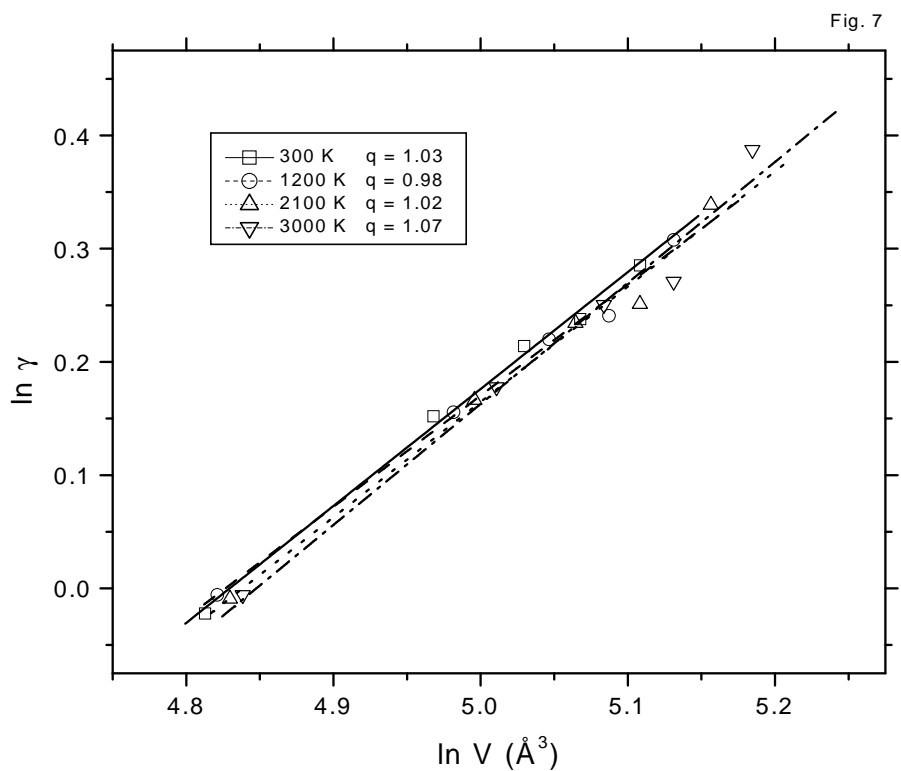,width=5in}}
\caption{The ln $\gamma$ versus ln V.  The slope of each curve
gives $q$: solid curve and squares, 300 K; dashed curve and circles,
1200 K; dotted curve and triangles, 2100 K; dash-dotted curve and
inverted triangles, 3000 K.  The lower temperatures are all close to
1, with a slight rise to 1.07 at 3000 K.}                  
\label{fig7}
\end{figure}

The volume dependence of $\gamma$, $q$, is 1.03 at 300 K
(\callout{Figure~\ref{fig7}}).
Our equation of state form, with $\Delta P_{\rm th}$ linear
in $T$ and independent of volume, constrains $q$ = 1 if
the isochoric heat capacity $C_{V} = 3 n R$, the
classical harmonic value.  As $T$ increases,
we find that
$q$ increases to a value of 1.07 at 3000 K
owing to anharmonicity.
Values from inverted experimental data sets range from
1.0 \cite{bina95} to 2.0 \cite{shim00}.  
{\it Stixrude et al.} [1992] found a high $q$ value
of 2.5 to go along with their value for $\gamma$.
{\it Akins and Ahrens{\rm 's}}
[1999] value was even higher, $q = 4.0 \pm 1.0$ with $C_{V} = 5nR$,
but these are preliminary conclusions based on
limited data.  {\it Patel et al.} [1996] found a value of 3.0
at 0 GPa, decreasing to 1.7 at 100 GPa using a combination
of molecular and lattice dynamics.  {\it Anderson{\rm 's}} [1998]
Debye calculations gave values close to unity at $T$ $\ge$ 1000 K,
with $q$ decreasing slightly with increasing $T$, to 0.82 at 1800 K.

\section{Discussion and Conclusions}
\begin{figure}
\centerline{\epsfig{file=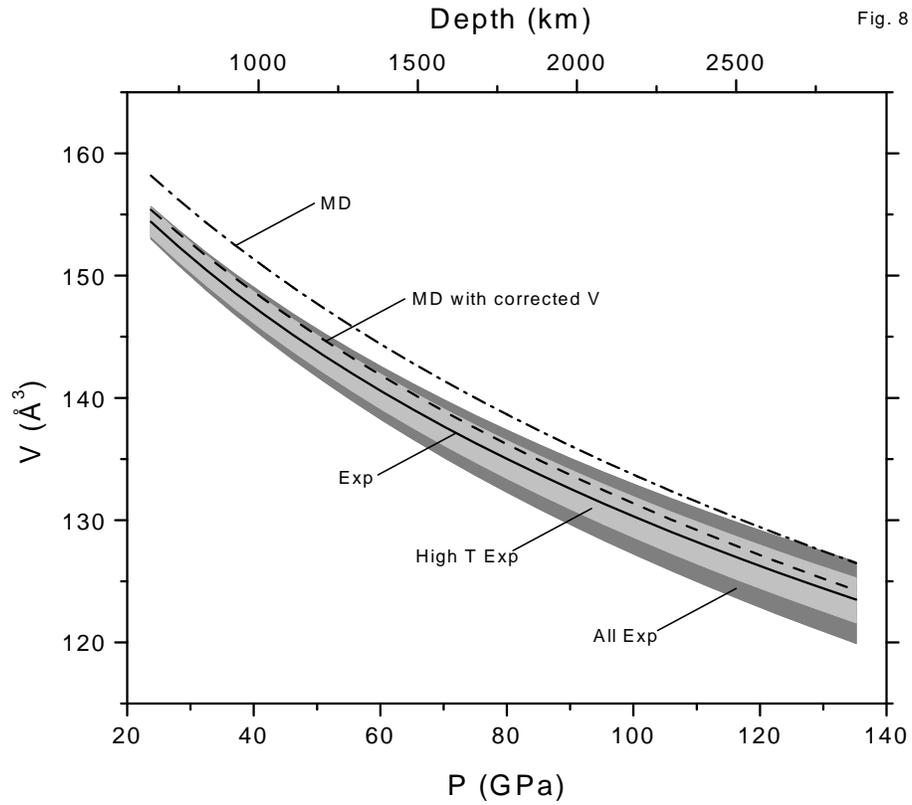,width=5in}}
\caption{Volumes of MgSiO$_{3}$ perovskite along the
geothermal gradient of {\it Brown and Shankland} [1981].
Dash-dotted curve is volume from our Universal EOS fit.  Dashed
curve is from our Universal fit with $V_{0}$ fixed to 162.49 \AA$^{3}$
[{\it Mao et al.,} 1991].  Dotted curve is for experimental
data sets with $V_{0}$ fixed to 162.49 \AA$^{3}$ (see Table 2).  
Both high temperature (F98+F00+S99) and all MgSiO$_{3}$ data
sets fall on the same line.
Shaded areas indicate the uncertainties in the fits of
the experimental data sets.}
\label{fig8}
\end{figure}

Taking sets of experimental P-V-T data, we fit the high
temperature data \cite{fiquet00, saxena99} and available experimental
MgSiO$_{3}$ data to our thermal Universal EOS form (Table 2)
in order to have consistent bases for comparison.
The resulting fits have higher statistical uncertainties
but compare well with our results.
Volumes calculated along the lower mantle geothermal gradient 
of {\it Brown and Shankland} [1981], which has a starting $T$ of 1873 K
at 670 km (Figure 1), 
show that those calculated from
our EOS are 3 to 4 \AA$^{3}$ per unit cell larger, $\sim$2.5\%,
than those calculated from the inverted experimental data sets
(\callout{Figure~\ref{fig8}}).  
This corresponds
to density differences of $\sim$0.1 g cm$^{-3}$.
This is the opposite of the typical error of the
local density approximation
to density functional theory, on which our method is based.
The larger volume is due to the choice of ionic charges
and other model assumptions.
If we fix $V_{0}$ to the experimentally determined value
of 162.49 \AA$^{3}$ \cite{mao91} but otherwise use our
EOS parameters, our calculated volumes are very close,
within 1\%,
to the volumes derived from the inverted experimental data sets.
The comparisons also improve as pressure increases.
To compare our model with other zero temperature theoretically derived 
EOSs, we calculated volumes at 0 K from 0-140 GPa and fit an 
isothermal Universal EOS to them (\callout{Table~\ref{tbl3}}).

Compositional models of the lower mantle can be tested
against seismological models by examining densities
and seismic wave velocities as functions of depth.
Candidates include pyrolite \cite{ringwood75}
and chondritic or pure perovskite models.
Studies support both the former \cite{bina90,stacey96}
and latter \cite{butler78,stixrude92}, though uncertainties
in the thermodynamic parameters of the constituent minerals
make it difficult to resolve this question with certainty.
Indeed, several studies have been able to support both
models depending on whether high (pure perovskite) 
or low (pyrolite) values of $\alpha$ and $\delta_{T}$ 
or $\gamma$ and $q$ are used \cite{zhao94,anderson95,karki99}.

\begin{figure}
\centerline{\epsfig{file=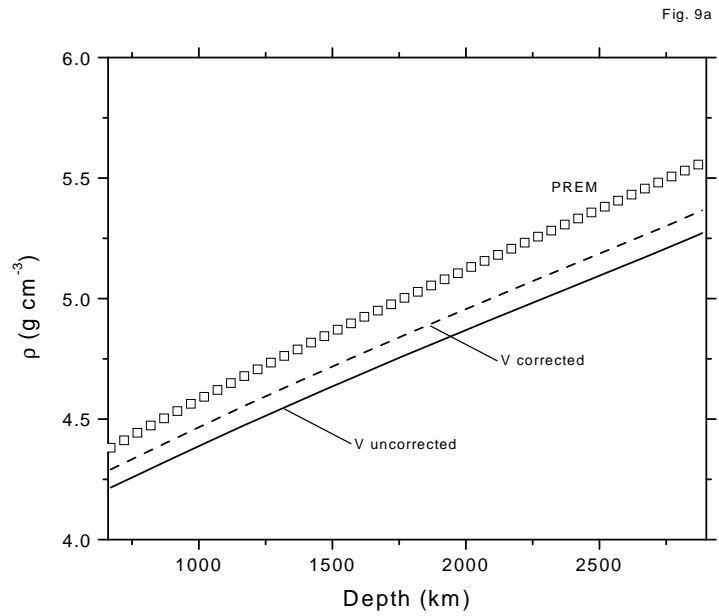,width=10cm}}
\centerline{\epsfig{file=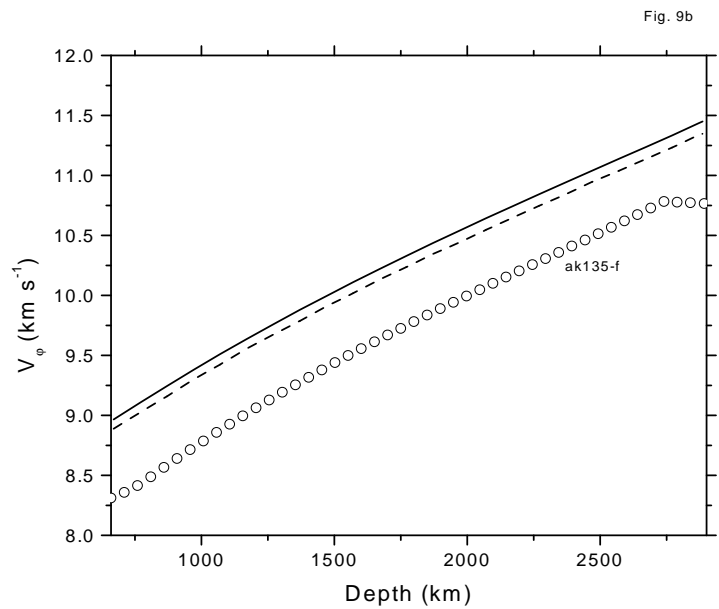,width=10cm}}
\caption{(a) Densities and (b) bulk sound velocities
of MgSiO$_{3}$ perovskite calculated
along the lower mantle geotherm of {\it Brown and Shankland} [1981]
using the thermal Universal EOS.
Solid curve is for uncorrected volumes and 
dashed curve is for $V_{0}$ set to the experimental value of
{\it Mao et al.,} [1991].
Squares are densities from PREM [{\it Dziewonski and Anderson,} 1981].
Circles are velocities from the ak135-f seismological model
[{\it Kennett et al.,} 1995; {\it Montagner and Kennett,} 1996].}
\label{fig9}
\end{figure}

Calculating the densities of Mg-pv along a
geothermal gradient \cite{brown81} (\callout{Figure~\ref{fig9}}a), 
we see that they are $\sim$0.2-0.3 g cm$^{-3}$ ($\sim$4-5\%)
too low compared with those from the
Preliminary Reference Earth Model (PREM) \cite{dziewonski81}.
The calculated bulk sound velocities,
$V_{\varphi} = \sqrt{K_{S} / \rho}$, where 
$K_{S} = K_{T} ( 1 + \gamma \alpha \, T )$,
are, in turn, $\sim$0.4-0.6 km s$^{-1}$ ($\sim$4-7\%) too high compared to
the ak135-f seismological model \cite{kennett95,montagner96} (Figure 9b).
Using the experimentally determined value of $V_{0}$ =
162.49 \AA$^{3}$ \cite{mao91} in place of our calculated
value does increase the densities somewhat, but not enough
to match the PREM values.  

\begin{figure}
\centerline{\epsfig{file=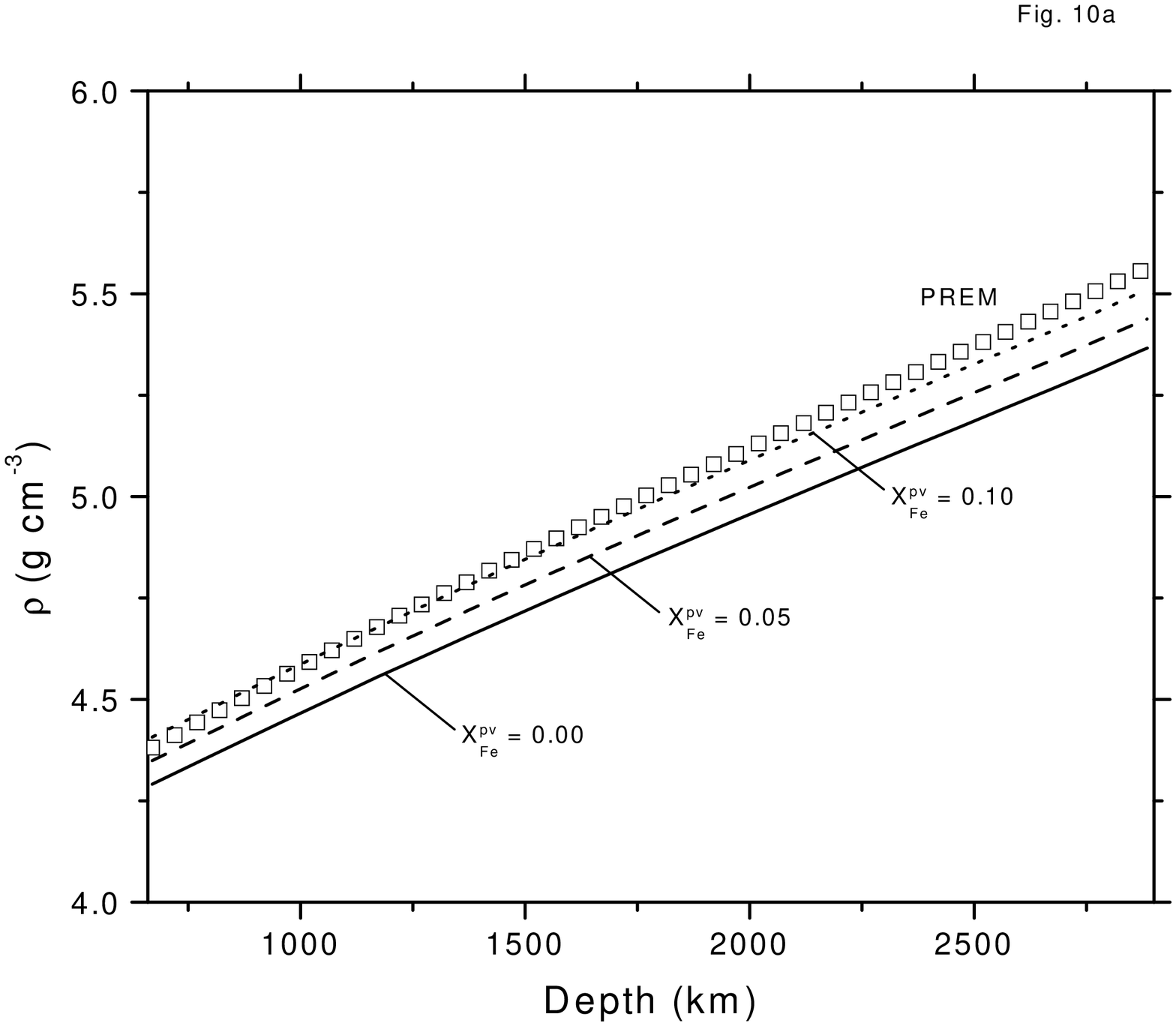,width=10cm}}
\centerline{\epsfig{file=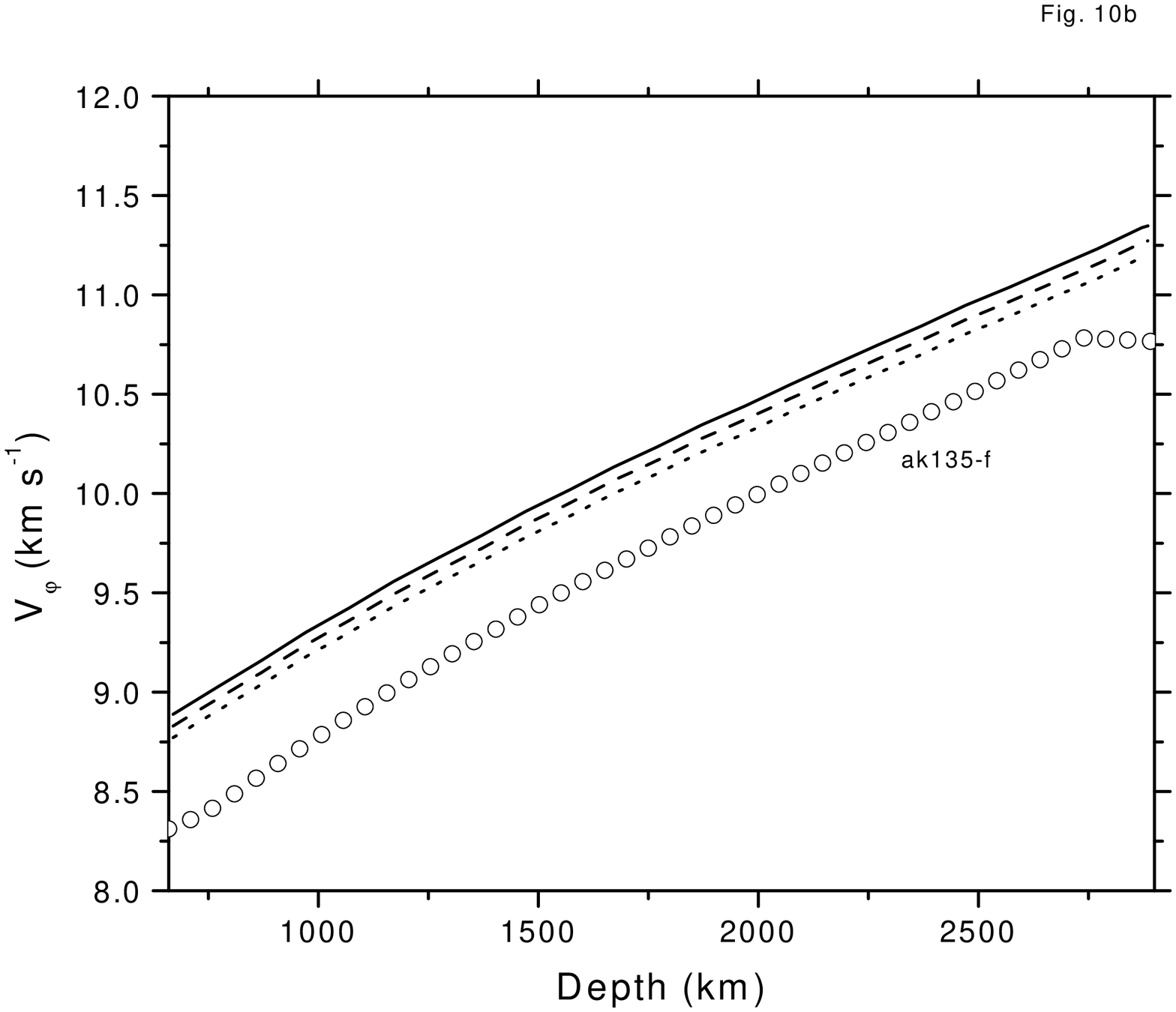,width=10cm}}
\caption{(a) Densities and (b) bulk sound velocities
of (Mg,Fe)SiO$_{3}$ perovskite calculated
along the geotherm of {\it Brown and Shankland} [1981]
using the experimental value of $V_{0}$
{\it Mao et al.,} [1991] for X${\rm ^{pv}_{Fe}}$ = 0.00 (solid curve),
0.05 (dashed curve), and 0.10 (dotted curve).
Squares are densities from PREM [{\it Dziewonski and Anderson,} 1981].
Circles are velocities from the ak135-f seismological model
[{\it Kennett et al.,} 1995; {\it Montagner and Kennett,} 1996].}
\label{fig10}
\end{figure}

As the perovskite found in
the lower mantle should be a solid solution of the Mg and 
Fe end-members, we added iron by adjusting density by
\cite{jeanloz83},
\begin{equation}
   \rho \, (X_{\rm {Fe}}) = \rho \, (0) (1 + 0.269 \, X_{\rm {Fe}})
\end{equation}
but we did not change the bulk moduli
\cite{mao91}.  Including
10 mol \% Fe does cause densities to agree somewhat with the
PREM values (\callout{Figure~\ref{fig10}}a), 
but the bulk sound velocities are still much too
high (Figure 10b).  Consequently, we find that the lower
mantle cannot consist solely of (Mg,Fe)SiO$_{3}$ perovskite.

\begin{figure}
\centerline{\epsfig{file=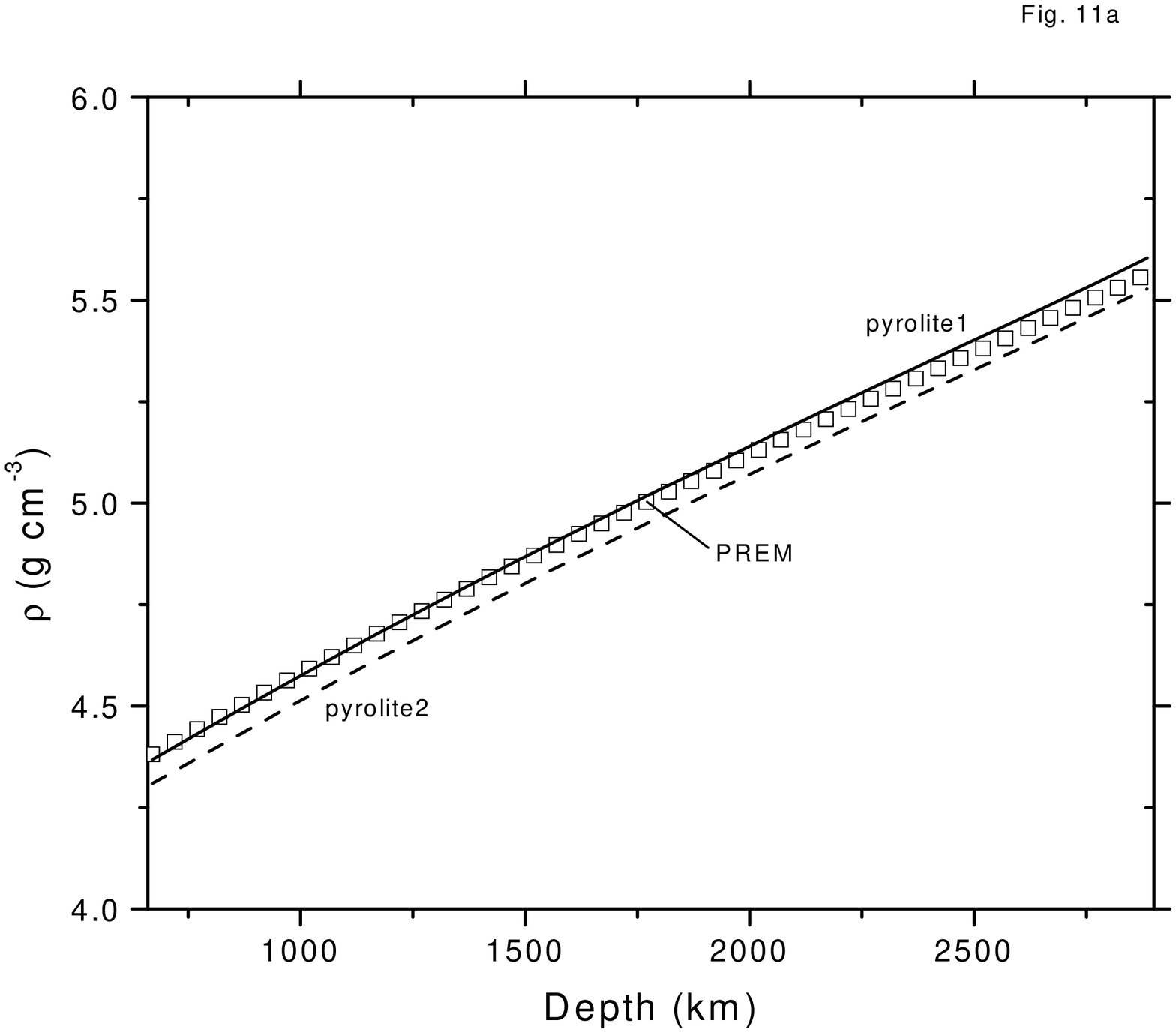,width=10cm}}
\centerline{\epsfig{file=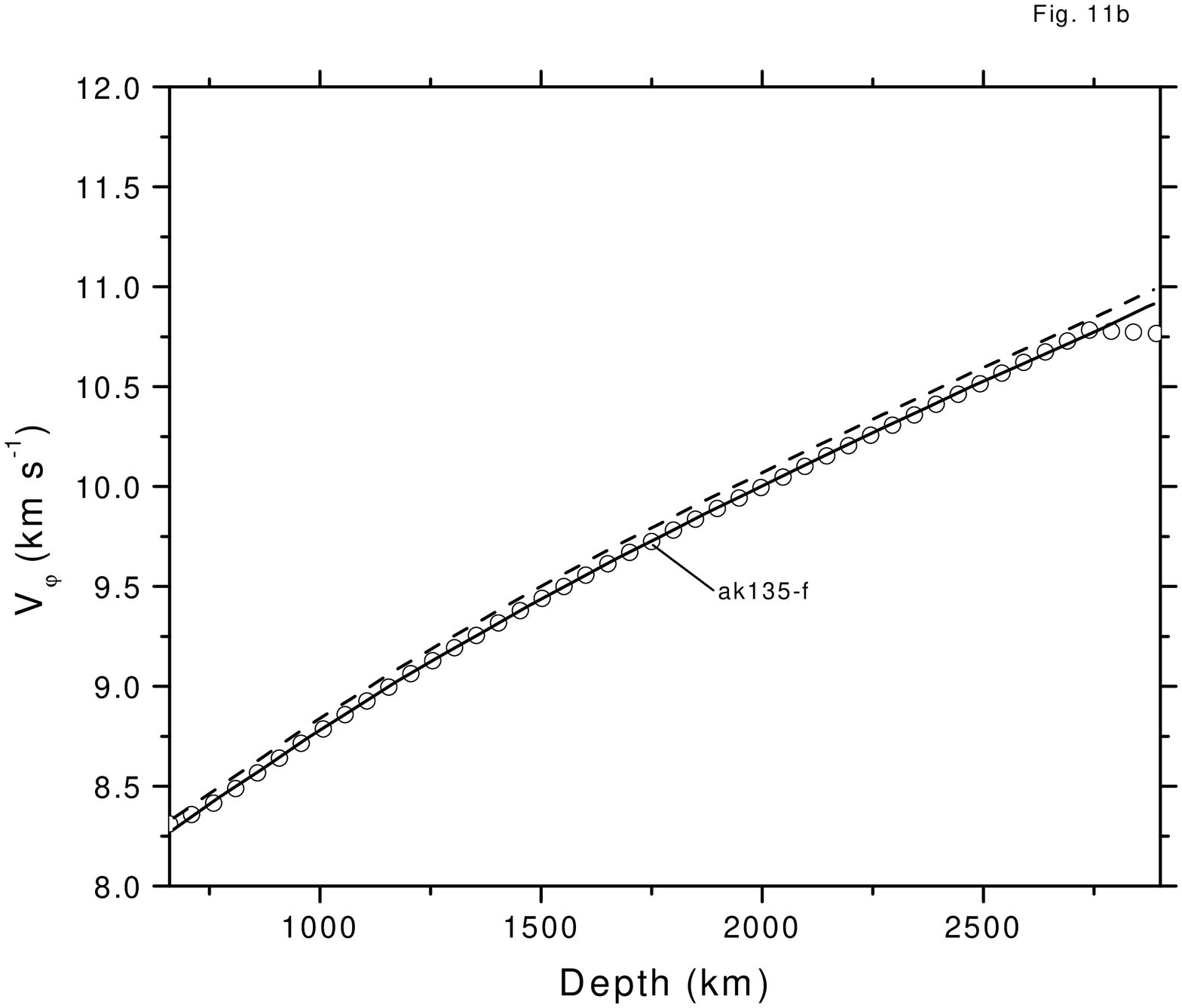,width=10cm}}
\caption{(a) Densities and (b) bulk sound velocities
of two simplified pyrolite models (67 vol \% pv, 33 vol \% mw)
along the geotherm of {\it Brown and Shankland} [1981].
Solid curve is X$^{\rm pv}_{\rm Mg}$ = 0.93, X$^{\rm mw}_{\rm Mg}$ = 0.82.
Dashed curve is X$^{\rm pv}_{\rm Mg}$ = 0.96, X$^{\rm mw}_{\rm Mg}$ = 0.86.
Squares are densities from PREM [{\it Dziewonski and Anderson,} 1981].
Circles are velocities from the ak135-f seismological model
[{\it Kennett et al.,} 1995; {\it Montagner and Kennett,} 1996].}
\label{fig11}
\end{figure}

We also tested pyrolitic compositions consisting of mixtures
of perovskite and magnesiow\"ustite (mw).  Densities of 
mw were calculated using the thermodynamic data set of 
{\it Fei et al.} [1991], as well as $K_{T}$s for the Mg
end-member.  $K_{S}$s were then calculated using the
relation 
\begin{equation}
   \alpha (P,T) = \alpha (P_{0},T) \left[
   \frac {V (P,T)}{V (P_{0},T)} \right] ^{\delta_{T}} ,
\end{equation}
with $\delta_{T}$ and $\gamma$ from {\it Inbar and Cohen}
[1995].  Iron was accounted for in $K_{S}$ via the
relationship \cite{jeanloz83}
\begin{equation}
   K_{S} \, (X_{\rm {Fe}}) = K_{S} \, (0) + 17 \, X_{\rm {Fe}} .
\end{equation}
X$_{\rm Mg}$ = 0.93-0.96 for perovskite (pv) and 0.82-0.86
for mw \cite{katsura96, martinez97} were used.
Densities and bulk sound velocities were calculated
using Voigt-Reuss-Hill averaging for
two simplified pyrolite models (high [1] and low [2] Fe content)
consisting of 67 vol \% pv and 33 vol \% mw and were found to be in 
excellent agreement with the seismological models
(\callout{Figure~\ref{fig11}}).
The two pyrolite models have partitioning coefficients
$K^{\rm {pv-mw}}_{\rm {Fe-Mg}}$ of 0.34 and 0.26, where
\begin{equation}
   K^{\rm {pv-mw}}_{\rm {Fe-Mg}} = 
   \frac {X^{\rm {pv}}_{\rm {Fe}}/X^{\rm {pv}}_{\rm Mg}}
   {X^{\rm {mw}}_{\rm {Fe}}/X^{\rm {mw}}_{\rm Mg}}.
\end{equation}
Experimental evidence suggests that for such a bulk
composition, $K^{\rm {pv-mw}}_{\rm {Fe-Mg}}$ should increase in the
mantle from $\sim$0.20 (660 km) to $\sim$0.35 (1500 km),
with X$^{\rm {pv}}_{\rm Mg}$ decreasing and 
X$^{\rm {mw}}_{\rm Mg}$ increasing
with depth \cite{mao97}.  Compositional
models with high Fe content of pv and low
Fe content of mw (and vice versa)
fall inbetween the results of those shown in Figure 11.
Given the range of Fe-Mg partitioning between perovskite
and magnesiow\"ustite under the appropriate pressure and
temperature conditions, we find pyrolite to be the most likely 
compositional model for the lower mantle.

Given that other components (Al, Fe$^{3+}$) and phases
(CaSiO$_{3}$ perovskite) should be present in the lower mantle,
we do not expect exact agreement of a simplified pyrolite
model with any seismological model.
It is believed that, under lower mantle conditions,
Al$_{2}$O$_{3}$ is mainly incorporated into the Mg-pv
structure \cite{irifune94,wood00}.  
Generally, the effect of the incorporation of Al into Mg-pv
on its physical properties has been considered small
[e.g., {\it Weidner and Wang,} 1998].  However, Al, unlike Fe, 
causes significant distortion in the Mg-pv lattice \cite{oneill94},
which may affect the compressibility.
Recent experimental work by {\it Zhang and Weidner} [1999],
at pressure of up to 10 GPa and temperatures of up to 1073 K,
indicates that compared with MgSiO$_{3}$, Mg-pv with 
5 mol \% Al has a smaller K$_{T}$ value
(232-236 GPa) and a $( \partial K_{T} / \partial T )_P$ value
more than double that in magnitude.  The values for $\alpha_{0}$,
$( \partial \alpha / \partial T )_P$, and $\delta_{T}$ are also larger.
Smaller bulk moduli and the higher density (0.2\% at 300 K and
0 GPa) would cause seismic velocities to decrease.
In addition, the presence of Al tends to equalize the partitioning
of Fe into perovskite and magnesiow\"ustite and
may allow garnet to coexist with perovskite in the uppermost
$\sim$100 km of the lower mantle \cite{wood96,wood00}.
These partitioning experiments, however, were done at 
high $f_{\rm{O}_{2}}$,
so it is uncertain how applicable they are to the mantle.
The effect of the presence of Fe$^{3+}$ in Mg-pv on the
EOS and elasticity, while known for defect concentrations and
electrostatic charge balance \cite{mccammon97}, is unknown.
As for CaSiO$_{3}$ perovskite (Ca-pv), 
experimental data suggest that its
elastic properties are in excellent agreement with
seismological models and thus would be invisible in the
lower mantle \cite{wang96,hama98.2}.

Performing MD calculations over the range pressures
and temperatures found in the Earth's mantle, we find
a thermal EOS that is in excellent agreement with experimental
results.  Thermodynamic parameters can be derived that agree well
with experimentally determined values and that can be
confidently interpolated to conditions found in the
lower mantle.  Moreover, no solid-solid phase transitions
or melting was found during the runs under lower mantle
conditions, so orthorhombic MgSiO$_{3}$ perovskite should 
be stable to the core-mantle boundary.
Using these results, we find that pyrolite
with $K^{\rm pv-mw}_{\rm Fe-Mg}$ = 0.26-0.34
is the most likely compositional model for the lower mantle.

\setcounter{equation}{0}

\acknowledgments
We thank Iris Inbar, Russell Hemley, Stephen Gramsch,
Joe Akins, Mark Woods, Sheila Coleman, and Lars Stixrude for 
helpful discussions and comments.
We acknowledge the support of the National Science
Foundation through grant EAR-9870328.
The computations were performed on the Cray SV1 at the Geophysical
Laboratory, with the support of NSF grant EAR-9975753 and the
W. M. Keck Foundation.

{}
\end{article}
   % You must type an \end{article}
   % command after the references.

\newpage

%%  FIGURE CAPTIONS
\clearpage

\clearpage

\clearpage

\clearpage

\clearpage

\clearpage

\clearpage

\clearpage

\clearpage

\clearpage

\clearpage

%%  TABLES

\begin{planotable}{rrcccc}
\tablewidth{41pc}
\tablecaption{MD P-V-T Data Used in EOS Fits}
\tablenum{1}
\tablehead{
   \multicolumn{1}{c}{P, GPa}&
   \multicolumn{1}{c}{T, K}&
   \multicolumn{1}{c}{a, \AA}&
   \multicolumn{1}{c}{b, \AA}&
   \multicolumn{1}{c}{c, \AA}&
   \multicolumn{1}{c}{V, \AA$^{3}$}
}
\label{tbl1}
\startdata
  0 &  300 & 4.8166 & 4.9283 & 6.9649 & 165.33 \nl
  0 &  600 & 4.8299 & 4.9397 & 6.9787 & 166.50 \nl
  0 &  900 & 4.8439 & 4.9519 & 6.9944 & 167.77 \nl
  0 & 1200 & 4.8586 & 4.9648 & 7.0107 & 169.11 \nl
  0 & 1500 & 4.8732 & 4.9769 & 7.0289 & 170.48 \nl
  0 & 1800 & 4.8886 & 4.9901 & 7.0455 & 171.87 \nl
  0 & 2100 & 4.9109 & 5.0145 & 7.0578 & 173.80 \nl
  0 & 2400 & 4.9311 & 5.0354 & 7.0688 & 175.52 \nl
  0 & 2700 & 4.9525 & 5.0538 & 7.0831 & 177.28 \nl
  0 & 3000 & 4.9672 & 5.0633 & 7.1076 & 178.76 \nl
 12 &  900 & 4.7790 & 4.8856 & 6.8895 & 160.86 \nl
 12 & 1200 & 4.7917 & 4.8951 & 6.9024 & 161.90 \nl
 12 & 1500 & 4.8036 & 4.9051 & 6.9165 & 162.97 \nl
 12 & 1800 & 4.8169 & 4.9157 & 6.9288 & 164.06 \nl
 12 & 2100 & 4.8295 & 4.9256 & 6.9441 & 165.18 \nl
 12 & 2400 & 4.8429 & 4.9367 & 6.9586 & 166.36 \nl
 12 & 2700 & 4.8628 & 4.9539 & 6.9690 & 167.88 \nl
 12 & 3000 & 4.9035 & 4.9531 & 6.9695 & 169.27 \nl
 25 &  300 & 4.6971 & 4.8093 & 6.7728 & 152.99 \nl
 25 &  600 & 4.7073 & 4.8169 & 6.7830 & 153.80 \nl
 25 &  900 & 4.7181 & 4.8250 & 6.7948 & 154.68 \nl
 25 & 1200 & 4.7285 & 4.8330 & 6.8055 & 155.53 \nl
 25 & 1500 & 4.7390 & 4.8409 & 6.8166 & 156.38 \nl
 25 & 1800 & 4.7497 & 4.8487 & 6.8291 & 157.27 \nl
 25 & 2100 & 4.7609 & 4.8572 & 6.8408 & 158.19 \nl
 25 & 2400 & 4.7728 & 4.8658 & 6.8514 & 159.11 \nl
 25 & 2700 & 4.7850 & 4.8736 & 6.8645 & 160.08 \nl
 25 & 3000 & 4.8015 & 4.8859 & 6.8773 & 161.34 \nl
 50 &  900 & 4.6194 & 4.7307 & 6.6445 & 145.20 \nl
 50 & 1200 & 4.6280 & 4.7366 & 6.6531 & 145.84 \nl
 50 & 1500 & 4.6366 & 4.7414 & 6.6632 & 146.48 \nl
 50 & 1800 & 4.6450 & 4.7474 & 6.6726 & 147.14 \nl
 50 & 2100 & 4.6545 & 4.7524 & 6.6821 & 147.81 \nl
 50 & 2400 & 4.6627 & 4.7589 & 6.6920 & 148.49 \nl
 50 & 2700 & 4.6728 & 4.7644 & 6.7014 & 149.20 \nl
 50 & 3000 & 4.6814 & 4.7700 & 6.7129 & 149.90 \nl
140 &  300 & 4.3572 & 4.5023 & 6.2698 & 123.00 \nl
140 &  600 & 4.3627 & 4.5046 & 6.2768 & 123.35 \nl
140 &  900 & 4.3687 & 4.5076 & 6.2839 & 123.74 \nl
140 & 1200 & 4.3747 & 4.5097 & 6.2903 & 124.10 \nl
140 & 1500 & 4.3806 & 4.5121 & 6.2967 & 124.46 \nl
140 & 1800 & 4.3867 & 4.5146 & 6.3027 & 124.82 \nl
140 & 2100 & 4.3923 & 4.5164 & 6.3104 & 125.18 \nl
140 & 2400 & 4.3986 & 4.5186 & 6.3170 & 125.55 \nl
140 & 2700 & 4.4048 & 4.5209 & 6.3235 & 125.92 \nl
140 & 3000 & 4.4109 & 4.5230 & 6.3309 & 126.31
\end{planotable}

\clearpage

\begin{planotable}{lllllll}
\tablewidth{41pc}
\tablecaption{Equation of State Parameters$^{\rm a}$}
\tablenum{2}
\tablehead{
   \multicolumn{1}{c}{}&
   \multicolumn{1}{c}{V$_{0}$, \AA$^{3}$}&
   \multicolumn{1}{c}{K$_{T_{0}}$, GPa}&
   \multicolumn{1}{c}{K$^{\prime}_{T_{0}}$}&
   \multicolumn{1}{c}{a, GPa}&
   \multicolumn{1}{c}{b, GPa K$^{-1}$}&
   \multicolumn{1}{c}{$\chi^{2}$, GPa}
}
\tablecomments{$^{\rm a}$MD are best fit EOS parameters to
molecular dynamics P-V-T data.  Experimental results
are the best fit Universal 
EOS parameters to experimental data sets, except data taken
from {\it Mao et al.} [1991] (M91). 
Remaining experimental results are 
F98, {\it Fiquet et al.} [1998] (27 data points);
F00, {\it Fiquet et al.} [2000] (38 data points); 
S99, {\it Saxena et al.} [1999], (37 data points);
All, all X$_{\rm Mg}$ = 1.0 data (363 data points; see Figure 1).
Values in parentheses are standard deviations.}
\tablecomments{$^{\rm b}$Fixed equal to 4.}
\tablecomments{$^{\rm c}$Fixed to $V_{0}$ from {\it Mao et al.} [1991].}
\tablecomments{$^{\rm d}$Fixed to $K_{T_0}$ from {\it Yeganeh-Haeri} [1994].}
\label{tbl2}
\startdata
This work \nl
$\:\:\:\:$Birch-Murnaghan EOS & 165.40(6)& 274(1) & 3.73(2) & -2.00(2) & 0.00667(5) & 0.08561 \nl
$\:\:\:\:$Universal EOS & 165.40(5) & 273(1) & 3.86(2) & -1.99(1) & 0.00664(5) & 0.06712 
\vspace{0.7em}\nl
Experimental results \nl
$\:\:\:\:$M91 & 162.49(7) & 261(4) & 4$^{\rm b}$ \nl
$\:\:\:\:$F98 & 162.65(15) & 262(6) & 3.41(22) & -1.79(21) & 0.00597(70) & 1.47438 \nl
& 162.49$^{\rm c}$ & 267(3) & 3.25(10) & -1.79(21) & 0.00597(70) & 1.41566 \nl
& 162.60(9) & 264$^{\rm d}$ & 3.33(2) & -1.79(21) & 0.00596(69) & 1.41333 \nl
& 162.49$^{\rm c}$ & $264^{\rm d}$ & 3.41(3) & -1.80(35) & 0.00601(73) & 1.36143 \nl
$\:\:\:\:$F00 & 162.74(56) & 239(8) & 4.41(18) & -1.85(6) & 0.00618(20) & 0.61052 \nl
& 162.49$^{\rm c}$ & 245(4)& 4.29(11) & -1.85(6) & 0.00618(20) & 0.59369 \nl
& 161.68(18) & 264$^{\rm d}$ & 3.88(2) & -1.85(6) & 0.00617(20) & 0.60570 \nl
& 162.49$^{\rm c}$ & $264^{\rm d}$ & 3.52(4) & -1.73(8) & 0.00577(28) & 1.23307 \nl
$\:\:\:\:$S99 & 163.25(63) & 242(6) & 4.45(10) & -1.72(1) & 0.00574(2) & 0.26719 \nl
& 162.49$^{\rm c}$ & 256(5) & 4.20(10) & -1.72(1) & 0.00572(2) & 0.26227 \nl
& 161.10(28) & 264$^{\rm d}$ & 4.06(4) & -1.71(1) & 0.00572(2) & 0.26645 \nl
& 162.49$^{\rm c}$ & $264^{\rm d}$ & 3.89(2) & -1.75(2) & 0.00584(7) & 0.31819 \nl
$\:\:\:\:$F98 + F00 + S99 & 162.93(33) & 235(6) & 4.72(13) & -1.85(12) & 0.00618(38) &
1.56307 \nl
& 162.49$^{\rm c}$ & 244(4) & 4.52(13) & -1.86(12) & 0.00621(40) & 1.55478 \nl
& 161.69(22) & 264$^{\rm d}$ & 4.08(1) & -1.87(12) & 0.00622(38) & 1.61901 \nl
& 162.49$^{\rm c}$ & $264^{\rm d}$ & 3.75(2) & -1.73(15) & 0.00577(7) & 2.22349 \nl
$\:\:\:\:$All & 162.31(10) & 249(6) & 4.40(20) & -1.92(10) & 0.00639(32) & 0.60337 \nl
& 162.49$^{\rm c}$ & 243(8) & 4.55(28) & -1.90(10) & 0.00635(34) & 0.61628 \nl
& 162.01(22) & 264$^{\rm d}$ & 3.92(8) & -1.88(10) & 0.00628(38) & 0.67860 \nl
& 162.49$^{\rm c}$ & $264^{\rm d}$ & 3.75(2) & -1.67(20) & 0.00557(65) & 1.04352
\end{planotable}

\vspace{3pc}

\begin{planotable}{llcl}
\tablewidth{21pc}
\tablecaption{Zero Temperature Equation of State Parameters}
\tablenum{3}
\tablehead{
   \multicolumn{1}{c}{}&
   \multicolumn{1}{c}{V, \AA$^{3}$}&
   \multicolumn{1}{c}{K, GPa}&
   \multicolumn{1}{c}{K$^{\prime}$}
}
\tablecomments{$^{\rm a}$Universal fit.  Other 
theoretical fits are third-order Birch-Murnaghan fits.}
\tablecomments{$^{\rm b}${\it Cohen} [1987].}
\tablecomments{$^{\rm c}${\it Stixrude and Cohen} [1993].}
\tablecomments{$^{\rm d}${\it Wentzcovitch et al.} [1995].}
\label{tbl3}
\startdata
This work$^{\rm a}$ & 164.22 & 280 & 3.84 \nl
PIB$^{\rm b}$ & 164.78 & 252 & 4.05 \nl
LAPW$^{\rm c}$ & 160.74 & 266 & 4.2 \nl
PWPP$^{\rm d}$ & 157.50 & 259 & 3.9
\end{planotable}

\clearpage

\end{document}